\documentclass[slac_one]{revtex4}
\usepackage{graphicx}
\usepackage{epsfig}
\usepackage{subfigure}
\usepackage{fancyhdr}
\begin{document}

\title{Performance of the ATLAS Minimum Bias Trigger in $p-p$ collisions at the LHC } 

%

\author{L. Tompkins on behalf of the ATLAS Collaboration}
\affiliation{UC Berkeley and LBNL, Berkeley, CA 94720, USA}

\begin{abstract}
The early physics program at the ATLAS experiment includes measuring the basic properties of proton proton collisions, such as charged particle multiplicities, in order to constrain phenomenological models of soft interactions in the LHC energy regime.  An inclusive and well understood trigger is crucial to minimize any possible bias in the event selection.  The ATLAS experiment uses two complementary types of minimum bias triggers.  A scintillator trigger sensitive to the forward regions of 2.1~$<~|\eta|~<~$3.8 has been proven to efficiently select proton proton collisions, while a trigger based on counting hits in the inner tracking detector has provided a useful control sample. The performance and efficiency measurements of these triggers and detectors will be presented. 
\end{abstract}

\maketitle

\thispagestyle{fancy}


\section{INTRODUCTION} 
\label{sec:intro}
The earliest measurements of the ATLAS experiment~\cite{det} at the LHC have been of the properties of charged particles in $p-p$ collisions~\cite{mb1, mb15}.  These measurements are critical for characterizing the low energy, non-perturbative regime of $p-p$ interactions, which are only be described with phenomenological models.  In order to produce measurements which are maximally inclusive and minimize model dependencies,  ATLAS has used a highly efficient single-armed scintillator trigger as the primary means of data collection.   A complementary trigger, based on randomly selected bunch crossings and information from the tracking detectors, has been used as a control trigger.  The excellent performance of these triggers have allowed ATLAS to measure charged particle properties at three center of mass energies within well defined phase spaces, including several with a minimum particle $p_{T}$ of 100~MeV~\cite{mb2,mb236}.  These proceedings document their performance.  Section~\ref{sec:det} describes the ATLAS detector and trigger system.  Section~\ref{sec:mbts} presents the scintillator detector performance and the minimum bias trigger efficiencies are shown in Section~\ref{sec:trigEff}.  Section~\ref{sec:conc} draws conclusions.  This work is further detailed in~\cite{mbtsconf}. 

\section{THE ATLAS DETECTOR AND MINIMUM BIAS TRIGGERS}
\label{sec:det}

The ATLAS experiment at the LHC is a multi-purpose detector designed for studying the full suite of physics measurements possible at a high energy proton collider.  The minimum bias triggers are provided by the Minimum Bias Trigger Scintillators (MBTS) and the silicon based tracking detectors, the Pixel detector and the Semi Conductor Tracker (SCT).  Triggering is facilitated by a three level architecture starting with hardware based triggers at Level 1. The individual sub detectors send Level 1 decisions to the Central Trigger Processor (CTP) which sets the global trigger bits.  Software decisions are made at Level 2 using partial detector information with the exception of the Inner Detector, for which all hits are read out.  The third level, the Event Filter, does a full event reconstruction in order to select the final events. 

The MBTS detectors consist of 2~cm thick polystyrene scintillators mounted 3.6~m from the nominal center of the detector.  They cover 2.09 $ < | \eta | < $ 3.84 and are divided into 2 rings in $|\eta|$ (2.09, 2.82) and (2.82,3.84), and 8 sections in $\phi$, allowing for 16 possible hits per detector side.  They are read out by wavelength shifting fibers embedded in the scintillator.  At Level 1 a leading edge discriminator determines if the energy deposited in a particular counter passes the threshold.  The outputs of all of the counters are sent to the CTP which sets the MBTS\_X bit, where X~$=$~1, 2 or 4, the minimum number of MBTS counters over threshold per event, or the MBTS\_X\_X bit, indicating at least X counters on each side of the the MBTS detector are over threshold.  The software based triggers were run in pass-through mode, although a more sophisticated fitting of the scintillator signals was used to compute the charge collected per counter at Level 2.

The Inner Detector provides charged particle tracking up to an $|\eta|$ of 2.5 and consists of three tracking technologies: two silicon based detectors and a third detector utilizing transition radiation in a straw tube structure.  The $mbSpTrk$ trigger uses the Pixel and SCT detectors.  The Pixel detector has 50x400~$\mu$m pixels arranged in three barrel layers and two endcaps of three layers each.  The SCT has 4 double sided layers of 80~$\mu$m wide silicon strips in the barrel and nine layers in each of two end caps.  Both systems have extremely low noise; the Pixel(SCT) detector has an occupancy of $10^{-9} (10^{-5})$ noise hits per event.  The $mbSpTrk$ trigger is fed by a random sampling of colliding bunches at Level 1.  At Level 2 clustering of the pixel and SCT hits is performed.  The pixel clusters consist of at least one pixel hit, while in the SCT at least one hit from each side of the double sided layers is required to form a cluster.  The $mbSpTrk$ trigger requires at least 4 pixel and 4 SCT clusters at Level 2.  The Event Filter has the ability to run track reconstruction to suppress beam backgrounds, however the background rate is negligible, therefore it is run in pass through mode.  
  
\section{MBTS PERFORMANCE FOR 7 TEV DATA TAKING}
\label{sec:mbts}
\begin{figure}[t]
\centering
\subfigure[]{\includegraphics[width=0.45\textwidth]{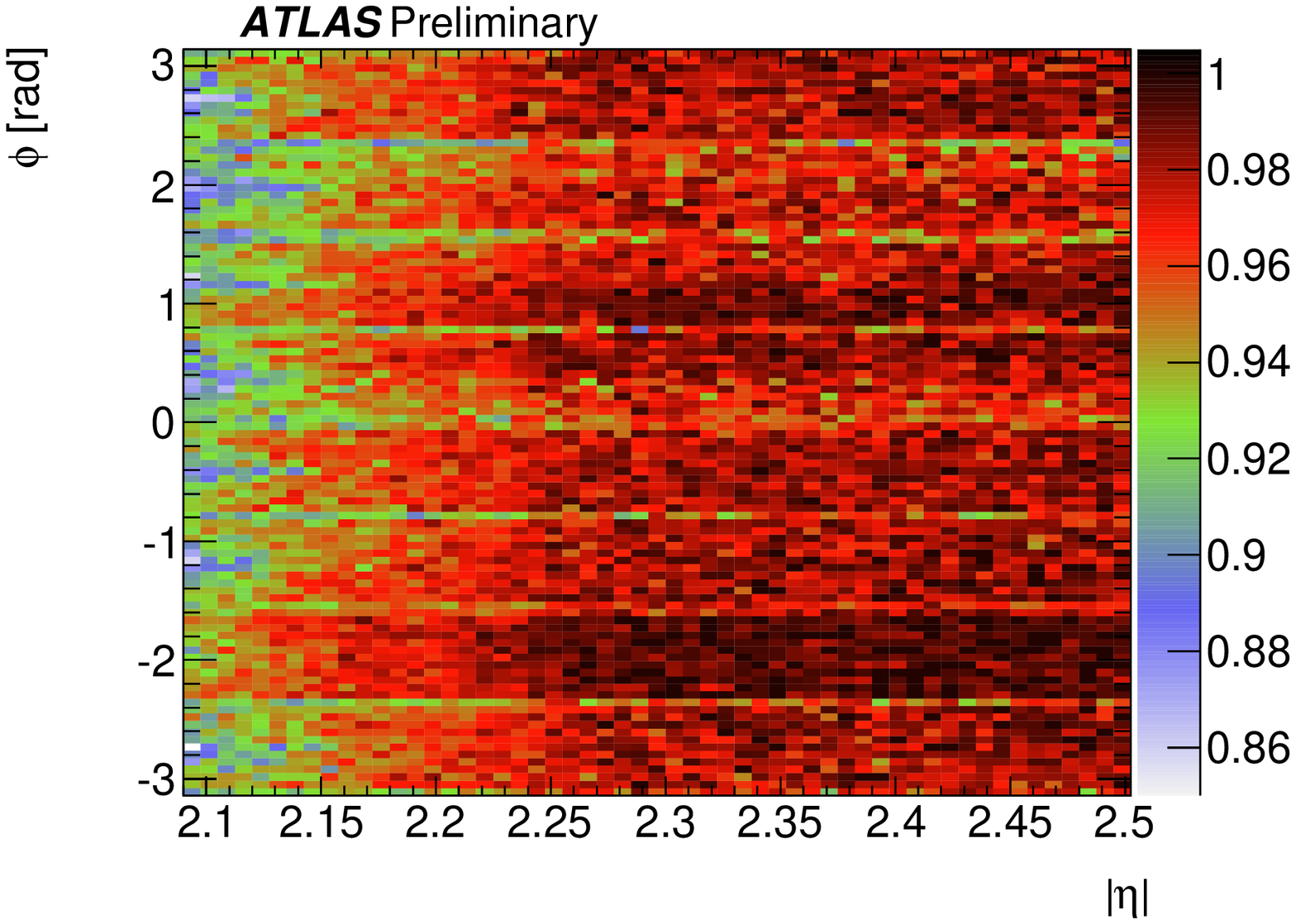}\label{fig:mbtsPerfa}}
\subfigure[]{\includegraphics[width=0.45\textwidth]{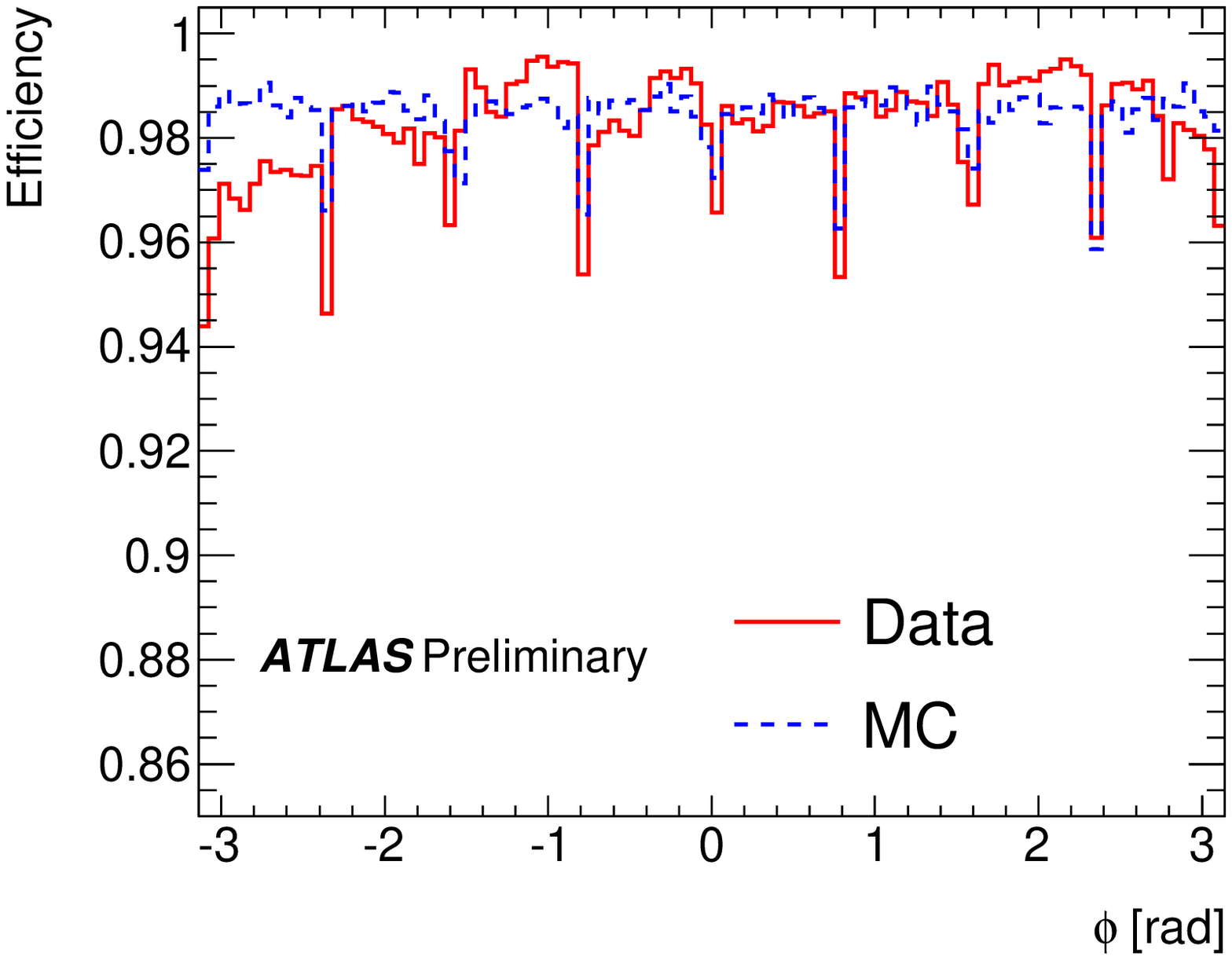}\label{fig:mbtsPerfb}}
\caption{\subref{fig:mbtsPerfa}: Efficiency for the A side of the MBTS counters measured in data as a function of the probe track $\eta$ and $\phi$ at the point of intercept with the MBTS. \subref{fig:mbtsPerfb}: Efficiency of the C side MBTS counters as a function of $\phi$ of the probe track.  In both cases see text for definition of the efficiency.} \label{fig:mbtsPerf}
\end{figure}

The MBTS response to charged particles was probed in data and Monte Carlo using tracks extrapolated to the detector in the region where the Inner Detector and MBTS overlap, 2.09~$<~|~\eta~|~<$~2.5.  Each track in this region was extrapolated to the MBTS and the resulting $\eta$ and $\phi$ position was used to determine which counter was hit.  An efficiency was defined by the fraction of times the intersected counter charge was greater than 0.15~pC, a threshold set to be above the noise level measured in data, when exactly one track intersected it.  Figure~\ref{fig:mbtsPerfa} shows this efficiency as a function of the $\eta$ and $\phi$ of the track in data.  It can be seen that the overall efficiency is high and that there are counter to counter variations.  The decrease in efficiency at low $\eta$ is dominated by extrapolation error on the tracks.  The right plot shows the efficiency integrated over 2.2 $<|\eta|<$2.5 for the C side of the MBTS compared for data and Monte Carlo.   Here the counter to counter variations of several percent are evident in the data whereas the MC shows a constant response.  The variations are attributed to differences in the operating points of the different counters.  Additionally the drop at the edges of the counters is more pronounced in the data due to the fact the the Monste Carlo does not model the decrease in light collection efficiency at the edges of the counters.

\section{MBTS TRIGGER EFFICIENCIES}
\label{sec:trigEff}

\begin{figure}[t]
\centering
\subfigure[]{\includegraphics[width=0.45\textwidth]{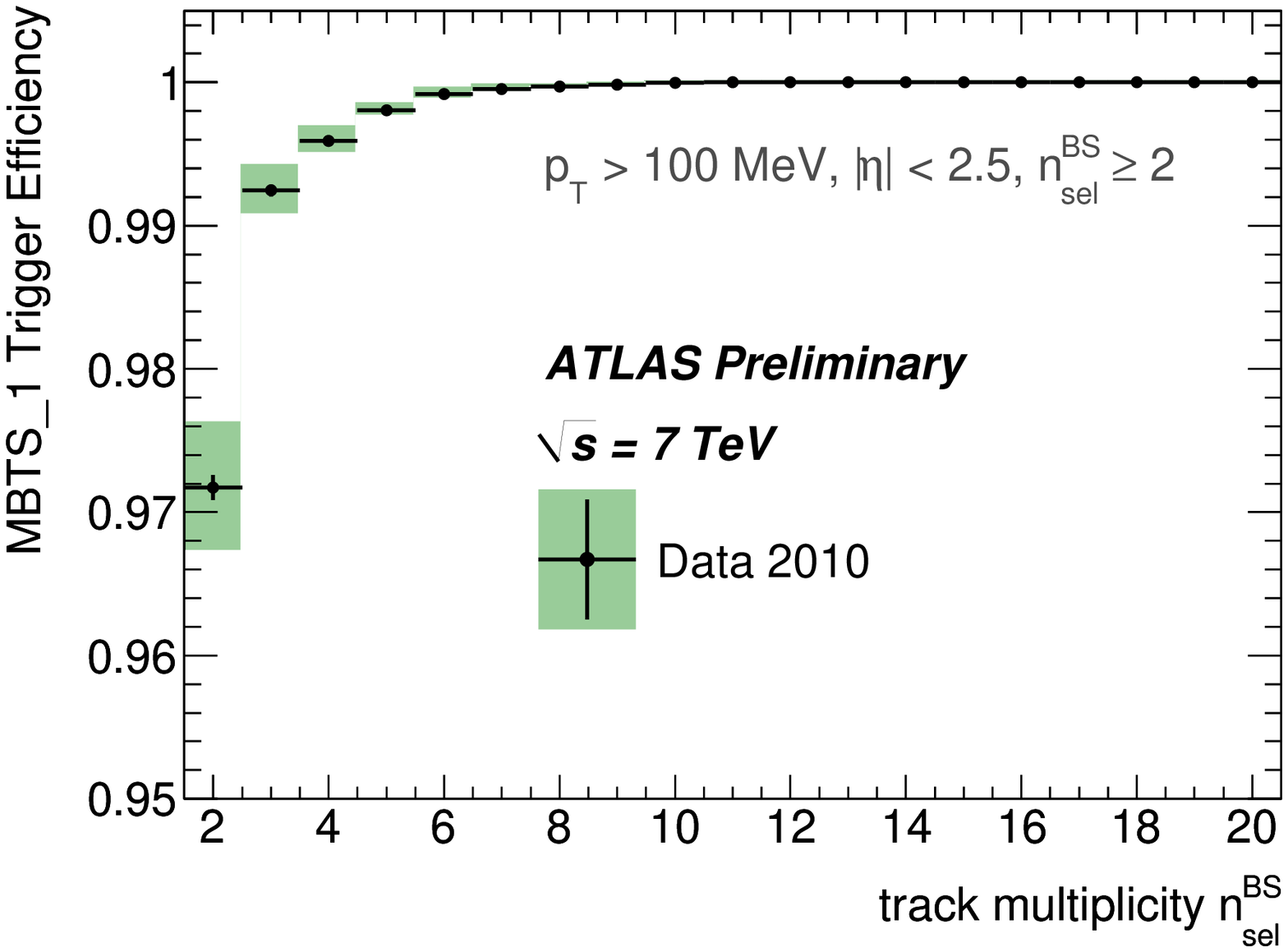}\label{fig:effa}}
\subfigure[]{\includegraphics[width=0.45\textwidth]{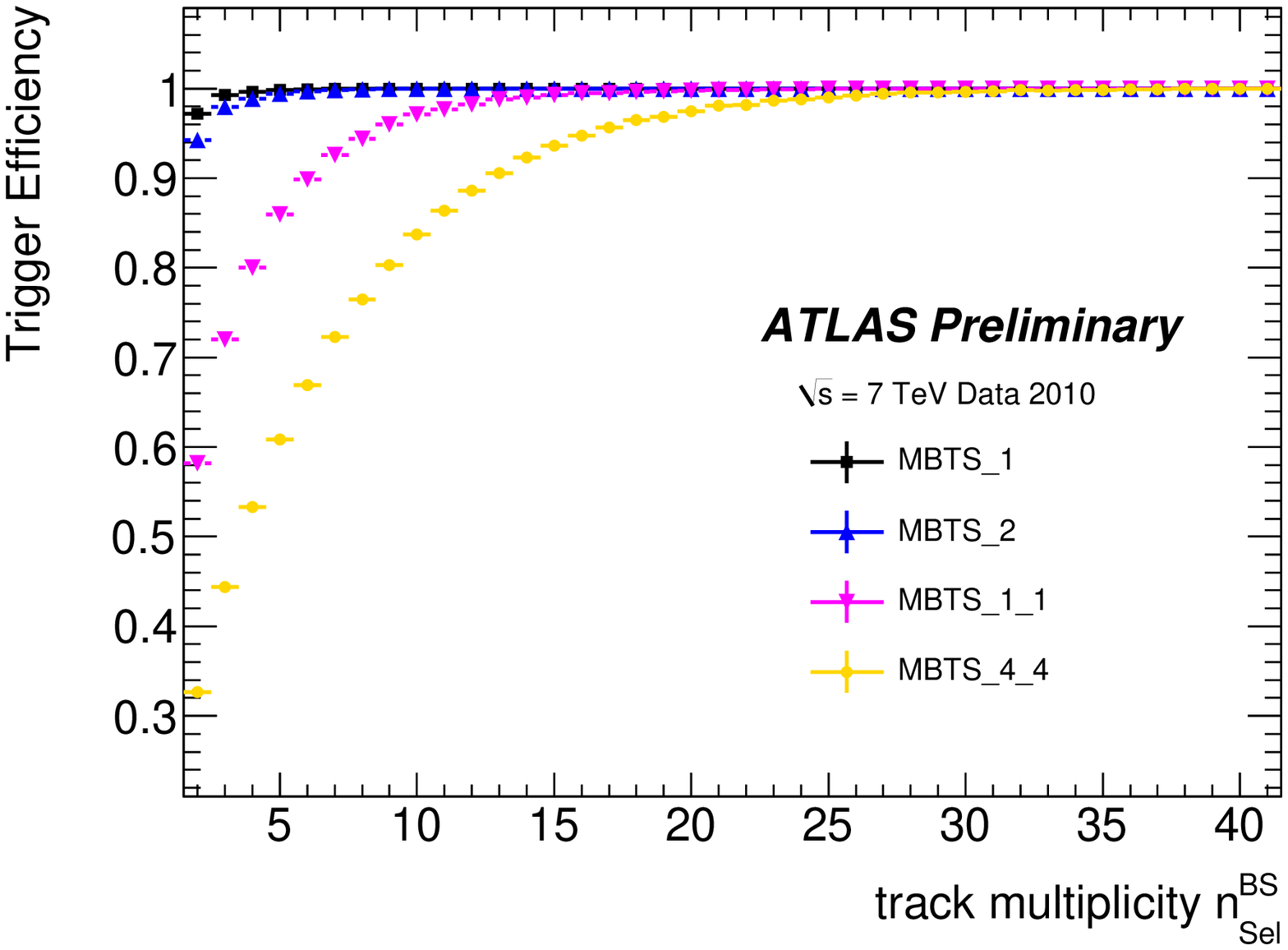}\label{fig:effb}}
\caption{\subref{fig:effa}: Efficiency of the MBTS\_1 trigger as a function of the number of selected tracks in the event with a minimum $p_{T}$ of 100~MeV.  \subref{fig:effb}:  A comparison of the MBTS\_1, MBTS\_2, MBTS\_1\_1 and MBTS\_4\_4 trigger efficiencies as a function of the number of tracks with $p_{T}  > $100~MeV in the event.} \label{fig:eff}
\end{figure}

The MBTS\_X trigger efficiencies were measured in data using events passing $mbSpTrk$ as the reference data sample.  They were measured as a function of the number of tracks in the event satisfying $p_T > $ 100~MeV, $|\eta| <2.5$, as well as quality cuts and impact parameter cuts with respect to the beam spot.  Both the beam background rates and the multiple interaction rates were negligible in the runs used for the measurement.  The efficiency is defined as $$\epsilon_{{\rm MBTS\_X}}  = \frac{{\rm MBTS\_X}~\&~mbSpTrk~\&~{\rm Offline Track}}{mbSpTrk~\&~{\rm Offline Track}}$$ and plotted as a function of the number of tracks in the event in Figure~\ref{fig:eff}.  Figure~\ref{fig:effa} shows that the MBTS\_1 trigger is highly efficient for events with low numbers of tracks, even with a low $p_{T}$ threshold.  The errors include systematic uncertainties due to the offline track selection and biases caused by using $mbSpTrk$ as the reference trigger.  The latter was investigated in Monte Carlo and found to be negligible.  Figure~\ref{fig:effb} shows the efficiencies for several MBTS trigger configurations.  Only statistical errors are shown.  It is noted that the MBTS\_2 trigger, which is less sensitive to beam related backgrounds, is nearly as efficient as the MBTS\_1 trigger.  Additionally the plots show that the requirement of coincidence, shown by MBTS\_1\_1 and MBTS\_4\_4, significantly reduces the efficiency for low multiplicities.  

\section{CONCLUSIONS}
\label{sec:conc}
The ATLAS Minimum Bias triggers have shown excellent performance during the early phase of the LHC physics program.  Because of their high efficiency and inclusiveness, these triggers have allowed ATLAS to measure charged particle properties of $p-p$ collisions with minimal model dependence and significant impact on the study of soft particle interactions.  




\begin{thebibliography}{9}   
\bibitem{det} ATLAS Collaboration, G.~Aad {\it et al.}, {\it The ATLAS Experiment at the Large Hadron Collider}, JINST 3 (2008) S08003.

\bibitem{mb1}
  ATLAS Collaboration,  {\it Charged-particle multiplicities in pp interactions at $\sqrt{s}$ = 900 GeV measured with the ATLAS detector at the LHC}, Phys. Lett. B~{\bf 688}, 21-42 (2010).
\bibitem{mb15} ATLAS Collaboration, {\it Charged particle multiplicities in pp interactions at $\sqrt{s}$ = 7 TeV measured with the ATLAS detector at the LHC}, ATL-CONF-2010-024.

\bibitem{mb2} ATLAS Collaboration, {\it Charged particle multiplicities in pp interactions for track $p_T$ $>$ 100 MeV at $\sqrt{s}$ $=$ 0.9 and 7 TeV measured with the ATLAS detector at the LHC}, ATL-CONF-2010-046.

\bibitem{mb236} ATLAS Collaboration, {\it Charged particle multiplicities in pp interactions at $\sqrt{s}$ $=$ 2.36 TeV measured with the ATLAS detector at the LHC}, ATL-CONF-2010-047.

\bibitem{mbtsconf} ATLAS Collaboration, {\it Performance of the Minimum Bias Trigger in p-p Collisions at $\sqrt{s}$ $=$ 7 TeV}, ATL-CONF-2010-068.

\end{thebibliography}
\end{document}